\documentclass[12pt]{article}

\def\ApJ{\sl Astrophys. J.}
\def\MNRAS{\sl Mon.Not.R. Astron. Soc.}

\title{\bf {Constraints On Galaxy Evolution Through Gravitational Lensing Statistics}}
\author{Deepak Jain\thanks{E--mail : deepak@ducos.ernet.in},
N. Panchapakesan\thanks{E--mail : panchu@vsnl.com},
S. Mahajan\thanks{E--mail : sm@ducos.ernet.in} and 
V. B. Bhatia\thanks{E--mail : vbb@ducos.ernet.in} \\
	{\em Department of Physics and Astrophysics} \\
        {\em University of Delhi, Delhi-110 007, India} 
	}

\usepackage{overcite}
\begin {document}
\maketitle

\begin{center}
\Large{\bf Abstract}
\end{center}
\large
\baselineskip=20pt

Explaining the formation and evolution of galaxies is one of the most
challenging problems in observational cosmology. Many observations
suggest that galaxies we see today could have evolved from the
merging of smaller subsystems. Evolution of galaxies tells us how the
mass or number density of the lens varies with cosmic time. Merging
between the galaxies and the infall of surrounding mass into galaxies
are two possible processes that can change the comoving number density
of galaxies and/or their mass. We consider five different evolutionary
models of galaxies .These models
are:  Non evolutionary model, Guiderdoni and Volmerange model, fast
merging, slow merging and mass accretion model. We study the
gravitational lens image  separation distribution function for these
models of evolving galaxies. 
A comparison with data for lensed quasars taken from the
HST Snapshot Survey rules out the fast merging model completely as this
model produces a large number of small-separation lenses. It is possible
that the mass 
accretion model and the non evolutionary model of galaxies may be able to
explain the small angle separations.

\begin {section} {Introduction}

After the discovery of the first multiply imaged quasars,
gravitational arcs and arc-lets, gravitational lensing has rapidly
become one of the most promising tools for cosmology. It is not a new
idea that the statistics of gravitational lensing can 
be used as a tool for the determination of cosmological
parameters.\cite{refs,press0}. In their pioneering
work, Turner, Ostriker \& Gott (1984, hereafter TOG)\cite{TOG} developed a 
formalism to calculate the lensing probability and image separation 
distributions. TOG modelled the lens population as point masses, singular
isothermal sphere (SIS) galaxies and cluster of galaxies. 
Since gravitational lens frequencies are sensitive to
the cosmological constant $( \Lambda)$, many authors
\cite{turn1,ft,f,mz,k96,chi} used the lens statistics, developed by
TOG, to constrain   
the cosmological parameters. Hinshaw \& Krauss \cite{hin} (1987) and
Krauss \& White \cite{kra}(1992) studied these effects with finite
core radii of galaxies. In all these  papers one
simplifying assumption is made that the comoving number density of
galaxies (lenses) is constant. However, it is an oversimplification to
assume that galaxies are formed at a single epoch. 

Mao\cite{m} (1991) first examined the effect of galaxy evolution on the
statistical properties of gravitational lenses using a simple redshift
cut-off model. This model reduces the lensing probability and also
explains the large mean separation of images in the observed gravitational
lenses. Sasaki \& Takahara \cite{st}(1993) used a more realistic model
than that 
of Mao and studied the effect of a more gradual evolution in the
number of galaxies on lens statistics. Mao \& Kochanek\cite{m1} (1994) put a
limit on  galaxy evolution by studying its effects on gravitational
lens statistics and image separations. They concluded that most of
the galaxies must have collapsed and formed by $z \sim 0.8$ if the
universe is  described by the Einstein-de Sitter model. If
elliptical galaxies are assembled from merging of spirals then most of
ellipticals must be formed by a redshift of 0.4. 

Rix et al. \cite{rix}(1994)  considered an evolutionary model which
is physically 
more plausible. They have studied the gravitational lens statistics with
an evolving lens population through the fast merging process only. They
showed that some specific merger models can be
rejected and $\Lambda$- dominated cosmologies are ruled out as they
predict a large number of sub-arc second lenses and with merging, this
problem  becomes
more acute. They find
that the lensing probability of getting multiple images is insensitive to
merging and the merging scenario skews $dN/d(\Delta\theta)$ towards smaller
separation.  Park \& Gott \cite{park}(1997) tried to explain  the
correlation 
between gravitational lens image separation and the source redshift in
the presence of the galaxy evolution. Jain et al.\cite{j1} (1998a)
also studied 
the effect of galaxy evolution on the statistical properties of lenses
both with  decaying and constant $\Lambda$.

The aim
of this paper is to use the image separation distribution function  of
lensed quasars, ($ dN/d{\Delta \theta}$), as a tool to put constraints
on the various evolutionary models of galaxies.
Hamana et al. \cite{ham}(1997) studied the distribution of image separation
angle of lensed quasars. However they found a complete mismatch between
the theoretical estimates and the observed lensed events. We find that
if we  use ($ dN/d{\Delta \theta}$) as a tool to study galaxy evolution,
it not only reduces the gap between observations and theoretical
predictions but also tells us which evolutionary model of galaxies
 might explain the observations. The paper is organized as follows. In $\S
2$ we explain the different evolutionary models of galaxies. In $\S
3$ we present the statistical formulas we use for comparing the model
calculations with the observations. We briefly summarize our results in
$\S 4$.

   \end {section}

\begin {section} {Evolution Of Galaxies}
\vskip 0.3cm
The theory of the formation and evolution of galaxies 
is one of the unsolved problems of astrophysics. 
Some authors believe that
galaxies evolve through a complex series of interactions before 
settling in the present day form \cite{too,schw}. Others believe that galaxies
were created in a well defined event at very early time\cite{egg,pa}. 
It remains
unclear which process dominates the formation of elliptical galaxies. 
Among the many theories of galaxy formation, the idea that galaxies may form by
the accumulation of smaller star forming subsystems has recently
received much attention. 
Many observations also support this
'bottom-up' scheme. 

First, deep Hubble Space Telescope (HST) images \cite{dri}
indicate that early type galaxies were assembled largely at $z >
1$ and have been evolving passively since $ z \leq 1$. Moreover, HST and
ground based telescopes show that the  galaxy-merger rate was higher
in the past and it roughly increases with redshift \cite{bur,carl}. This 
suggests that the galaxies we see today could have been assembled from
the merging of smaller systems sometime before $z \sim 1$.

Recent observations\cite{z} also show
that elliptical galaxies are rarer at high redshifts than those
predicted by 
models in which elliptical galaxies completed their star formation by
$ z \sim 5$. Therefore elliptical galaxies must have had significant
star formation at $ z < 5$ through merging and associated
starbursts. The formation of elliptical galaxies in this way is also
consisitent with the predictions of hierarchical clustering models of
galaxy formation.

The second piece of evidence comes from the excess of faint blue
galaxies which has been found in many deep imaging studies
\cite{gla,lil,eil}. Comparison with the model which assumes that no
luminosity evolution takes place in 
the galaxy population, shows that in the B band the actual observed
galaxy count 
exceeds the model predictions by a factor of 5. Merging of galaxies can
solve the surprisingly steep increase in the number density of
galaxies
\cite{rocca,gui,bro,car}. But at present it is not clear whether such
models adequately 
describe the merging of galaxies in any realistic models of structure
formation in the universe.

We consider five different evolutionary models of galaxies which try to
explain some of the observational facts listed above . These models
are:  Non evolutionary model, Volmerange and Guiderdoni model, fast
merging, slow merging and mass accretion model. We study the
gravitational lens image  separation distribution function in the
presence of evolving models of galaxies. 
\eject

\noindent {\bf{\underline{Non- Evolutionary Model}}}
\vskip  0.5cm
\noindent This is the conventional lens model in which the luminosity
function 
of lens galaxies is assumed to be of the  Schechter form \cite{sch}

\begin{equation}
\Phi(L,z = 0) =
\phi^\ast (L /L_\ast)^\alpha exp(- L/ L_\ast) dL/L_\ast
\end{equation}

\noindent where the $\phi^{*}$, $\alpha$ and $L^{*}$ are the
normalization factor, the index of faint - end slope, and the
characteristic luminosity respectively. These values are fixed in order
to fit the current luminosities and densities of galaxies.
This model assumes that the comoving number density of galaxies $n(\delta t)$
is constant and the mass of galaxies does not change with  cosmic time.

\begin{equation}
n(\delta t) =  n_{0} = constant
\end{equation}

\noindent where $\delta $t is the look-back time. The velocity 
dispersion of
Singular Isothermal Sphere (SIS) lenses at  $\delta $t is

\begin{equation}
v(\delta t) =  v_{0} = constant
\end{equation}

The subscript $0$ refers to  present-day values.

\vskip 0.5cm
\noindent{\bf{\underline{Volmerange and  Guiderdoni Model}}}
\vskip 0.5cm
In 1990, Volmerange and Guiderdoni \cite{rocca}, proposed a unifying
model to 
explain faint galaxy counts as well as observational properties of 
distant radio galaxies. This model of galaxy evolution is based on
number evolution in addition to pure luminosity evolution. According
to this model the present day
galaxies result from the merging of a large number of building blocks
and the comoving number of these building blocks evolves as $ ( 1 +
z)^{1.5}$.

It is argued that the present luminosity function is the
well known Schecter Luminosity Function \cite{sch} given in eq.(1)
above. Then at
high z, the comoving number density follows New Luminosity
Function 
\begin{equation}
\Phi(L,z)dL = ( 1 + z)^{2\eta} \Phi(L(1 + z)^\eta,0) dL 
\end{equation}
 It is seen that the value $\eta = 1.5$ gives a fair fit to
the data on high redshift galaxies. The functional form has the 
following properties:
\begin{enumerate}
\item [(i)] Self-similarity as suggested by the  Press-Schecter
(1974)\cite{press1} 
formalism subject to the constraint that the total mass of
associated material is conserved. 

\item [(ii)] The comoving number density evolves as
$\phi^\ast(z)$=$\phi^\ast_{0}$ $(1 + z)^\eta$ and the  characterstic
luminosity of the self similar galaxy luminosity function varies
as $L_\ast(z) = L_\ast(0)(1 + z)^{-\eta}$
\end{enumerate}

\vskip 0.5cm
\noindent {\bf{\underline{Fast Merging Model}}}
\vskip 0.5cm
The first merger model is that of
Broadhurst, Ellis \& Glazebrook (1992) \cite{bro}, which was 
originally motivated
by the faint galaxy population counts. This model assumes the number
density of the lenses to be a function of the look back time $\delta 
t$ as:
  
\begin{equation}
n(\delta t) = f(\delta t) n_{0}
\end{equation}

\noindent The velocity 
dispersion of
SIS  lenses at  $\delta $t is

\begin{equation}
v(\delta t) = [f(\delta t)]^{-1} v_{0}
\end{equation}

\noindent This form implies that if we had $n$ galaxies at  time
$\delta t$ each with velocity dispersion $v$, they would by today have
merged into one galaxy with a velocity dispersion $[f(\delta t)]v
$. The strength and the time dependence of merging is described by the
function $f(\delta t)$:

\begin{equation}
f(\delta t) = exp(Q H_{0}\delta t)
\end{equation}

where $H_{0}$ is the Hubble constant at the present epoch and Q represents 
the merging 
rate. We take Q = 4 \cite{bro}. The 
look back time $\delta t$ is related to the redshift  $z$ through

\begin{equation}
H_{0}\delta t = \int^{z}_{0}{(1+y)^{-1} dy \over{\sqrt{F(y)}}}
\end{equation}
where  $F(y)=\Omega_{0}(1 + y)^{3}+(1-\Omega_{0}-\Lambda_{0})(1 +
y)^{2}+\Lambda_{0}$ , 

\noindent $\Omega_{0} = 8 \pi G \rho_{0} /3 H_{0}^2$ , $\rho_{0}$ is
the density of matter, $\Lambda_{0} = 8 \pi G \rho_{v0} /3 H_{0}^2$
and $\rho_{v0}$ is density
of vacuum energy.

\vskip 0.5cm
\noindent{\bf{\underline {Slow Merging Model}}}

\vskip 0.5cm
\noindent In this less extreme merger model the total mass of 
galaxies within a given comoving volume is conserved. The comoving
number density goes like $t^{-2/3}$ while the mass of an individual
galaxy increases like $t^{2/3}$, where $t$ is the cosmic time since
the big bang\cite{gun,park}. We further assume the mass-velocity
 relation $ M \propto v^{\gamma} $. The values of $\gamma$ and
$v_{*}$ are given in Table 2 for elliptical galaxies. Then

\begin{equation}
n(\delta t) =  n_{0}\left[1 - {\delta t \over t_{0}}\right]^{-2/3}
\end{equation}

\begin{equation}
v(\delta t) = v_{0}\left[1 - {\delta t \over t_{0}}\right]^{2\over 3\gamma}
\end{equation}

\noindent where $t_{0}$ is present age of the universe.

\vskip 0.5cm

\noindent{\bf{\underline{Mass Accretion Model}}}

\vskip 0.5cm

\noindent Mass accretion is the key factor for evolution of
galaxies. A galaxy  can accrete mass through two processes: either it
accretes gas regularly through internal dynamics or the accretion
occurs in more violent events, galaxy interactions and
mergers. This is in line with the idea of hierachical formation.
In this model the comoving density of the galaxies is  
constant but the mass increases as $t^{2/3}$ as in the cosmological
infall model \cite{park}. The total mass in galaxies thus
increases with time. The comoving number density and the dispersion
velocity  vary as

\begin{equation}
n(\delta t) =  n_{0}(constant)
\end{equation}

\begin{equation}
v(\delta t) = v_{0}\left[1 - {\delta t \over t_{0}}\right]^{2\over 3\gamma}
\end{equation}

   \end {section}

\begin{section} {Basic Equations For Gravitational Lensing Statistics}

\noindent
The differential probability $d\tau$ of a beam encountering a lens 
in traversing the path of $dz_{L}$ is given by

\begin{equation} d\tau = n_{L}(z)\sigma{cdt\over dz_{L}}
dz_{L},
\end{equation}.

\noindent
 where $n_{L}(z)$ is 
the comoving number density [TOG 1984].

The Singular Isothermal Sphere (SIS) provides us with a reasonable
 approximation to 
account for the lensing properties of a real galaxy. The lens model
 is characterized
by the one dimensional velocity dispersion $v$. The deflection
 angle for all impact parameters is given by $\tilde{\alpha} =
 4\pi v^{2}/c^{2}$. The lens produces two images if the angular
 position of the source is less than the critical angle $\beta_{cr}$,
which is the deflection of a beam passing at any radius through an SIS:

\begin{equation} \beta_{cr} =\tilde{\alpha} D_{LS}/D_{OS} ,\end{equation}

\noindent we use the notation $D_{OL} =
d(0,z_{L}),D_{LS} = d(z_{L},z_{S}), D_{OS} = d(0,z_{S})$, where
$d(z_{1},z_{2})$ is the angular diameter distance between the redshift
$z_{1}$ and $z_{2}$ \cite{f}. 
\noindent
Then the critical impact parameter is defined by $a_{cr} = 
D_{OL}\beta_{cr}$ and the cross- section is given by 

\begin{equation}
\sigma = \pi a_{cr}^{2} = 16{\pi}^{3}\left({v \over
c}\right)^{4}\left({D_{OL}D_{LS}\over D_{OS}}\right)^{2}  ,
\end{equation}

\noindent
\vskip 0.3in

{\bf {\underline{ The Evolutionary Model}}}
\vskip 0.5cm
The differential probability $d\tau$ of a lensing event in an
evolutionary model can be written as\cite{j2,j3}:

\eject
\begin{eqnarray}
{d\tau}&=& {16\pi^{3}\over{c
H_{0}^{3}}}\,\phi_\ast\, v_\ast^{4}\Gamma\left(\alpha + {4\over\gamma} +1\right)f(\delta t)^{( 1 - {4\over\gamma})} \nonumber\\
& & \nonumber\\
&&\times\,(1 + z_{L})^{3}
\left({D_{OL}D_{LS}\over R_\circ D_{OS}}\right)^{2}{1\over
R_\circ}{cdt \over dz_{L}} dz_{L}\end{eqnarray}  

$${d\tau} = F(1 + z_{L})^{3}\left({D_{OL}D_{LS}\over R_\circ
D_{OS}}\right)^{2} f(\delta t)^{( 1 - {4\over\gamma})} 
{1\over R_\circ}{cdt \over dz_{L}} dz_{L}$$.

where  $ F ={16\pi^{3}\over{c
H_{0}^{3}}}\,\phi_\ast\, v_\ast^{4}\Gamma\left(\alpha + {4\over\gamma}
+1\right)$. The functional form of  $f(\delta t)$ in various models is
described in Table 1.
\vskip 0.2cm
\noindent
\begin{table}[ht]
\begin {center}
Table 1
\end{center}
\begin{center}
The Functional Form of $f(\delta t)$ 
\end{center}

\begin{center}
\begin{tabular}{|l|l|}\hline\hline
$Evolutionary\,\, Model$ & $ Form\,\, of\,\, (f\delta t)  $ 
 \\   \hline
\hline
$Fast\,\, Merging$ & $exp(Q H_{0}\delta t)$  \\ 

$VG\,\, Model$ & $(1 + z)^{1.5}$ \\

$Slow \,\,Merging$ & $\left(1 - {\delta t \over t_{0}}\right)^{-2/3}$    \\

$Mass \,\,Accretion ^{a}$ &$\left(1 - {\delta t \over t_{0}}\right)^{-2/3}$    \\

\hline
\end{tabular}

\end{center}
superscript 'a': In this case the  exponent of $
f(\delta t)$  in eq. (16) becomes 
 $(-1 -{ 4\over \gamma})$ as the total mass in galaxies increases with time.
\end{table}

Using the relations $ \phi = {\frac{\Delta\theta}{8 \pi
(v^{*}/c)^2}}$, ${\Delta\theta} = 8 \pi
(\frac{v}{c})^2{\frac{D_{LS}}{D_{OS}}}$ and luminosity 
\vskip 0.2cm
- velocity
relation $ {\frac{L}{L_{*}}} = ({\frac{v}{v_{*}}})^{\gamma}$, we get
${\frac{L}{L_{*}}} = ({D_{OS}\over{D_{LS}}} \phi)
^{\frac{\gamma}{2}}$. $\Delta\theta$ is image angular separation.
The differential optical depth of lensing in traversing
$dz_{L}$ with angular separation between $\phi$ and $\phi + d\phi$ is
given by

\begin{eqnarray}
{\frac{ d^{2} \tau }{ dz_{L}d\phi}}{d\phi}{dz_{L}}
&=& F(1+z)^{3} 
exp \left[- ({D_{OS}\over{D_{LS}}} \phi) ^{\frac{\gamma}{2}}f(\delta
t) \right] \nonumber \\
& &\times{(\frac{D_{OL} D_{LS}}{ D_{OS} R_{\circ}})^2 { 1\over R_{\circ}} {c
dt\over dz_{L}}f(\delta t)^{(2 + \alpha)}} \nonumber \\
&   &\times({D_{OS}\over{D_{LS}}} \phi) ^{{\gamma\over 2}(\alpha +
1+{4\over \gamma}) }({\gamma/2 \over{\Gamma(\alpha + 1+
{4\over\gamma})}}){\frac {d\phi}{\phi}}{dz_{L}} 
\end{eqnarray}

In eq.(17) the
exponent of $ f(\delta t)$ becomes 
$\alpha$ for the mass accretion model while in other evolutionary
models the exponent of 
$f(\delta t)$  remains $\alpha + 2 $.
After integrating eq.(17) over the lens redshift $z_{L}$, we
obtain the 
angular separation distribution  

\begin{equation}
{\frac{dN}{d\Delta\theta}} = <B> \sum \int_{0}^{z_{s}}{\frac{ d^{2}
\tau }{dz_{L}d\Delta\theta}} {dz_{L}}
\end{equation}

where the summation is over sources in the HST quasar sample and $< B>$
is the averaged bias which is equal to 9.76 in the HST quasar
sample\cite{ham}.
\noindent

\vskip .3in

{\bf {\underline{ The Non Evolutionary Model}}}
\vskip 0.5cm
\noindent
In the non- merging model the optical depth is given by \cite{f}

\begin{eqnarray}
{d\tau} &=& {16\pi^{3}\over{c
H_{0}^{3}}}\,\phi_\ast\, v_\ast^{4}\Gamma\left(\alpha + {4\over\gamma}
+1\right)(1 + z_{L})^{3} \nonumber\\ 
& & \nonumber\\
&&\times\,
\left({D_{OL}D_{LS}\over R_\circ D_{OS}}\right)^{2}{1\over
R_\circ}{cdt \over dz_{L}} dz_{L}\end{eqnarray}  

It is clear from eq. (16) that if $\gamma = 4$ then optical depth at
each redshift in evolutionary model of galaxies (except mass accretion
model) becomes equal to the optical depth in the non evolving model of
galaxies. The differential optical depth of lensing in traversing
$dz_{L}$ for angular separation between $\phi$ and $\phi + d\phi$ for the
non evolutionary model is given by

\begin{eqnarray}
{\frac{ d^{2} \tau }{ dz_{L}d\phi}}{d\phi}{dz_{L}}
&=& F(1+z)^{3} 
exp \left[- ({D_{OS}\over{D_{LS}}} \phi) ^{\frac{\gamma}{2}} \right]
\nonumber \\
& &\times{(\frac{D_{OL} D_{LS}}{ D_{OS} R_{\circ}})^2 { 1\over R_{\circ}} {c
dt\over dz_{L}}} \nonumber \\
&   &\times({D_{OS}\over{D_{LS}}} \phi) ^{{\gamma\over 2}(\alpha +
1+{4\over \gamma}) }({\gamma/2 \over{\Gamma(\alpha + 1+
{4\over\gamma})}}){\frac {d\phi}{\phi}}{dz_{L}} 
\end{eqnarray}

The $dN/d(\Delta\theta)$ which is given by eq.(18), strongly depends
upon the four  parameters 
$\alpha$, $\gamma$, $\phi^{*}$ and dispersion velocity $v_{*}$. We
consider three sets of these parameters given in Table 2.

\begin{table}[ht]
\begin {center}
Table 2
\end{center}
\begin {center}
Schechter and Lens Parameters for E/S0 Galaxies
\end{center}

\begin{center}
\begin{tabular}{|l|lllll|}\hline\hline
$Survey$ & $ \alpha$ & $ \gamma$ & $ v^{*} (Km /s)$ & $\phi^{*}
(Mpc^{-3})$ & $ F^{*}$  
 \\   \hline
\hline
$K96$ & $- 1.0$ &  $4.0$ &$225$ & $6.1 \times 10^{-3}$& $0.026$ \\ 

$LPEM$ & $+ 0.2$ & $4.0$ &$205$ &$3.2 \times 10^{-3}$& $0.010$   \\

$NS97$ & $- 1.1$ & $3.3$ & $175$ &$1.1\times 10^{-2}$& $ 0.015$   \\
\hline
\end{tabular}

\end{center}

References: K96 - C. S. Kochanek (1996)\cite{k96}.
            LPEM - J. Loveday et al. (1992)\cite{lov}.
	    NS97 - T.T. Nakamura \& Y. Suto (1997) \cite{naka}

\end{table}

\end {section}


\begin{section} {Result and Discussion}

An obvious noticeable fact from eq. (16) is that when the index $\gamma$
(the Faber-Jackson index) is different from the value of four that
we get any dependance on the evolution of galaxies. The combination
$(1-4/\gamma )$ vanishes when $\gamma$ has the value 4 and totally
suppresses the effect of evolution.

Figs. 1 to 3 give the calculated and observed number of lensed objects
as a function of the image separation for the currently accepted values of
cosmological parameters \cite{per,ad} $\Omega = 0.3$ and $\Lambda = 0.7$.
In all
these figures the histograms indicate the image separation distribution
of the four lensed quasars observed in the HST snapshot survey.
 It is clear
from these figures that there is a very significant dependance of the
results on the
Schecter and lens parameters. The uncertainties
associated with these parameters \cite{chen} will be diminished
greatly by the 
data that emerges from the surveys planned for the next few years.
The K96 set of parameters shift the peak of
the image separation distribution function ( $dN/d(\Delta \theta)$)
for all the
evolving models towards smaller angles of separations. The larger angle 
separations are not explained by these parameters. With LPEM  
parameters, the mass accretion model and, to a lesser extent, the 
non-evolution model seem to have reasonable success in explaining the 
observations.
Figs 4 to 6 present the results for the case of zero cosmological 
constant. Here the best fit seems to be the K96 parameters.

It is clear that to reach at a firm conclusion about galaxy evolution
more reliable and valid set of Schechter parameters  
are badly needed. This should be forthcoming in the next few years with
the large number of surveys that are in progress and that are being 
planned for the future.

Satisfactory and unambiguous identification of lensed objects should also
improve in the coming years. The X-ray observations by CXO (Chandra 
X-ray Observatory) should be specially helpful.\cite{mun}. 
 An increase in the overall number of lensed objects at all 
wave lengths is also expected in the next few years. It is hoped that 
knowledge of cosmological parameters and galaxy distribution (Schechter)
parameters will mutually refine each other as more results 
become available. It will then be possible to say with some certainty
which of the various galaxy evolution models are favoured by
observations and which are ruled out.

   \end {section}

\begin{section}*{Acknowledgements}
We thank E. Turner, D. Maoz, Takashi Hamana and Yu- Chung N. Cheng
for useful discussions.

   \end {section}

\vfill
\eject

\begin{figure}[ht]
\vskip 15 truecm
\includegraphics{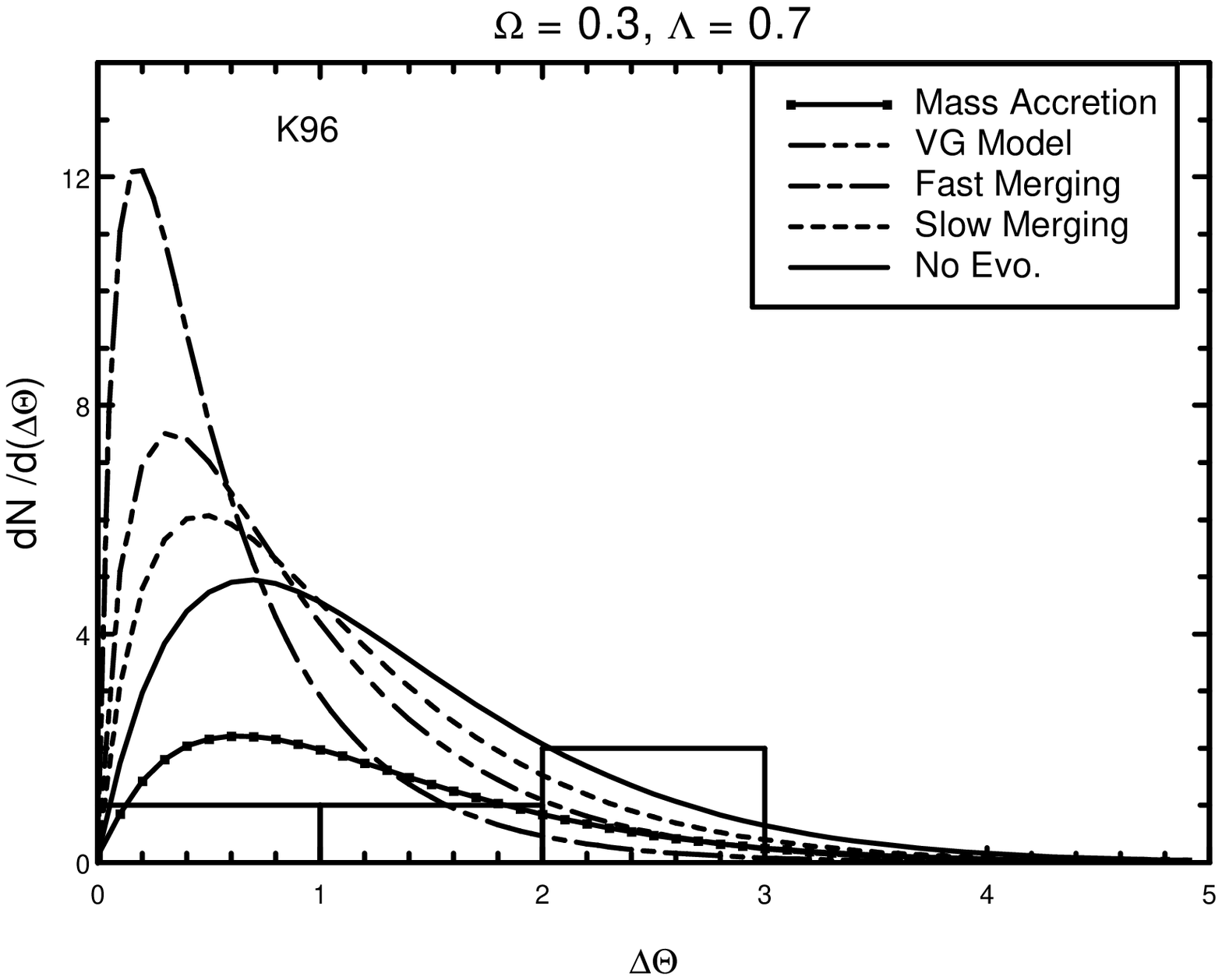}
\caption{}{ The expected distribution of lens image separations
with K96 parameters with $\Omega = 0.3, \Lambda = 0.7$}
\end{figure}

\begin{figure}[ht]

\vskip 15 truecm
\includegraphics{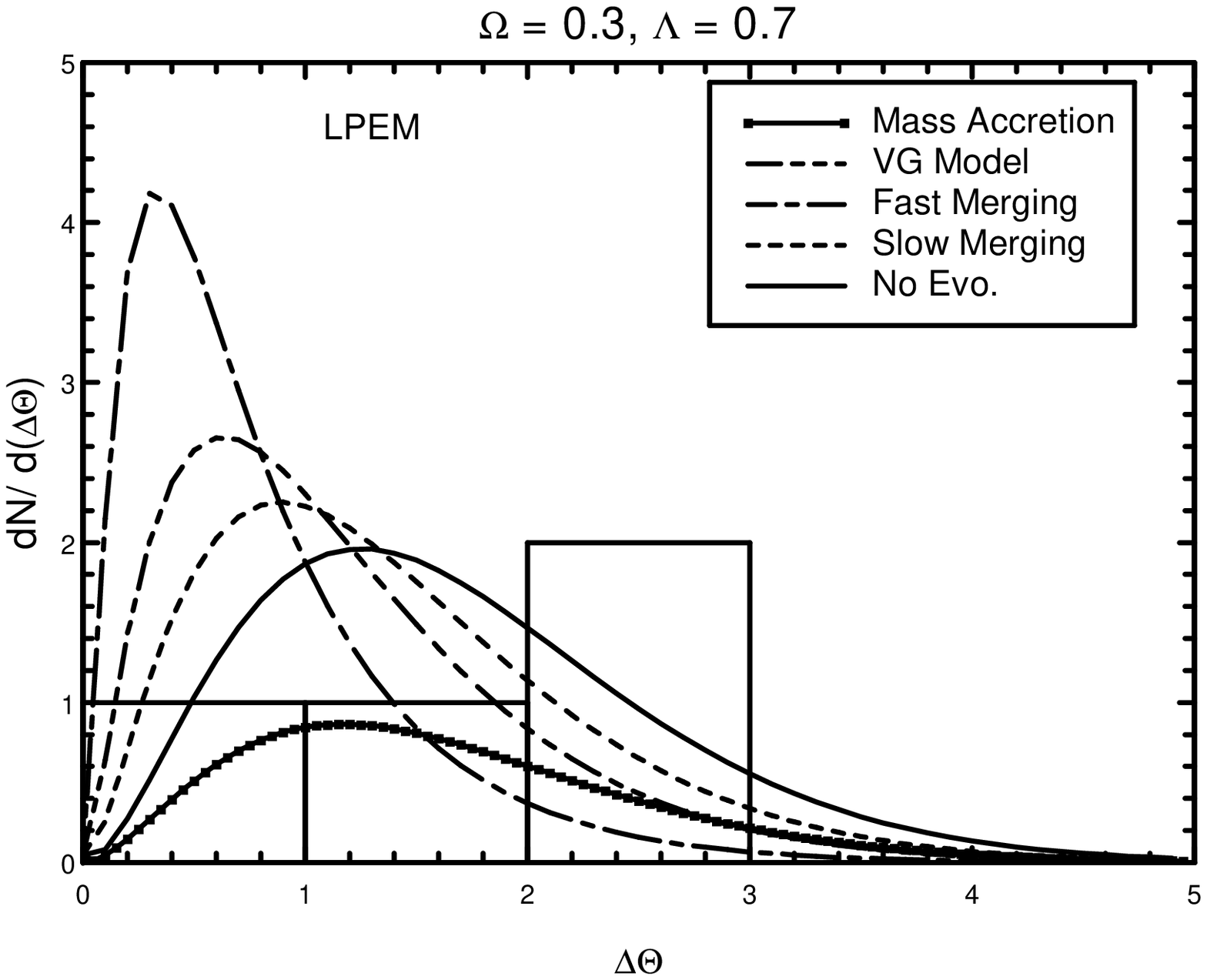}
\caption{}{ The expected distribution of lens image separations
with LPEM parameters with $\Omega = 0.3, \Lambda = 0.7$}
\end{figure}

\begin{figure}[ht]
\vskip 15 truecm
\includegraphics{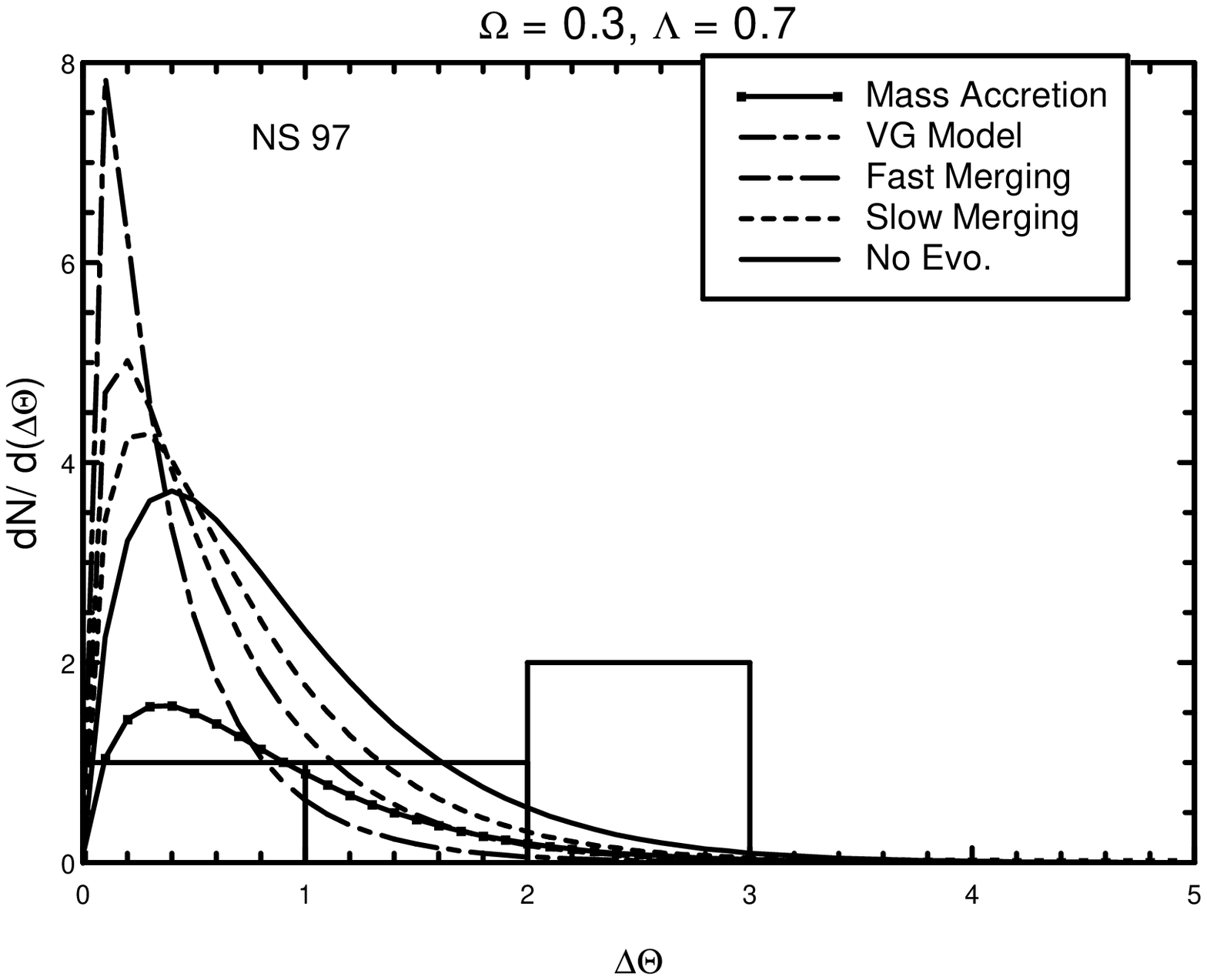}
\caption{}{ The expected distribution of lens image separations
with NS97 parameters with $\Omega = 0.3, \Lambda = 0.7$}
\end{figure}

\begin{figure}[ht]
\vskip 15 truecm
\includegraphics{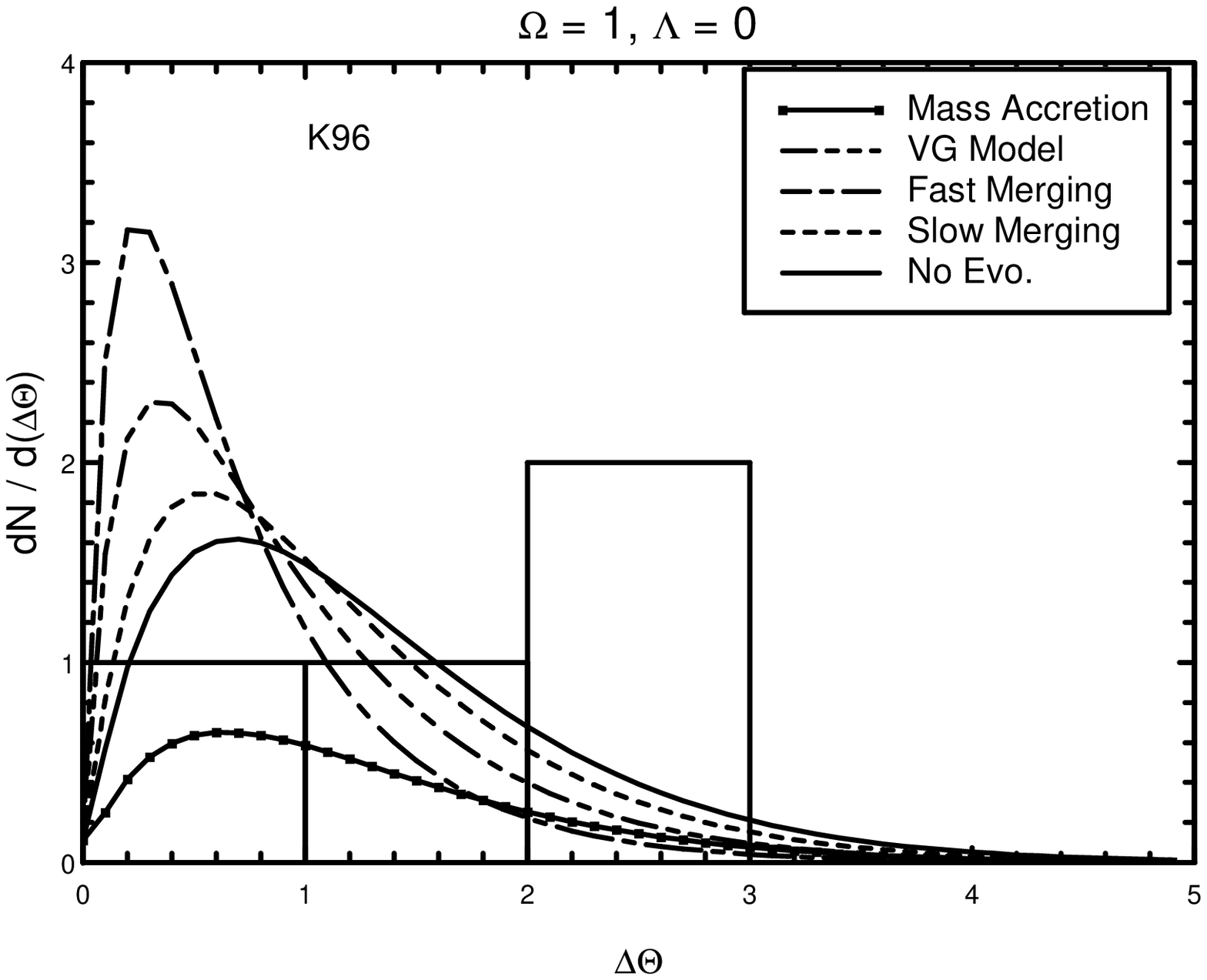}
\caption{}{ The expected distribution of lens image separations
with K96 parameters with $\Omega = 1.0, \Lambda = 0.0$}
\end{figure}

\begin{figure}[ht]
\vskip 15 truecm
\includegraphics{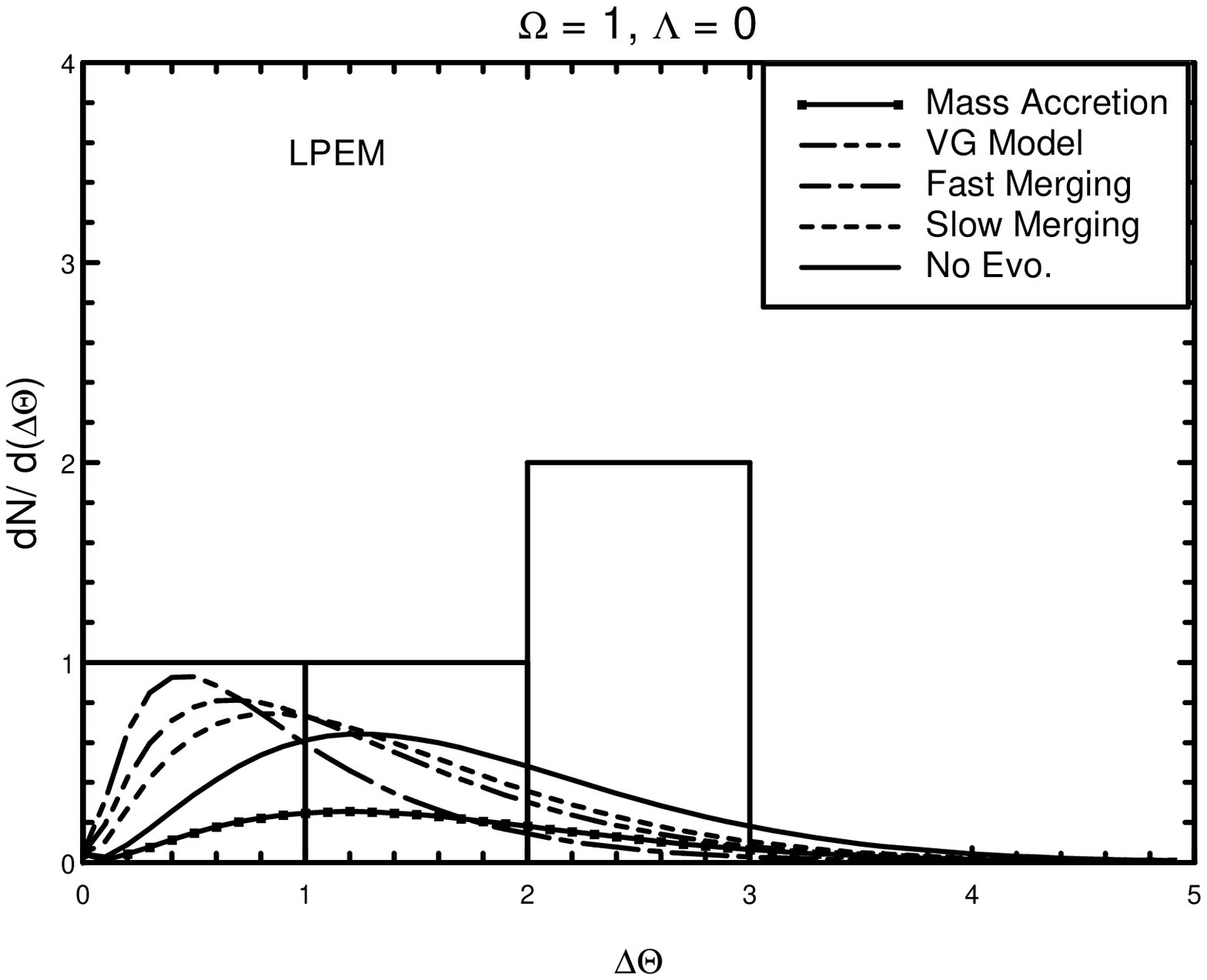}
\caption{}{ The expected distribution of lens image separations
with LPEM parameters with $\Omega = 1.0, \Lambda = 0.0$}
\end{figure}

\begin{figure}[ht]
\vskip 15 truecm
\includegraphics{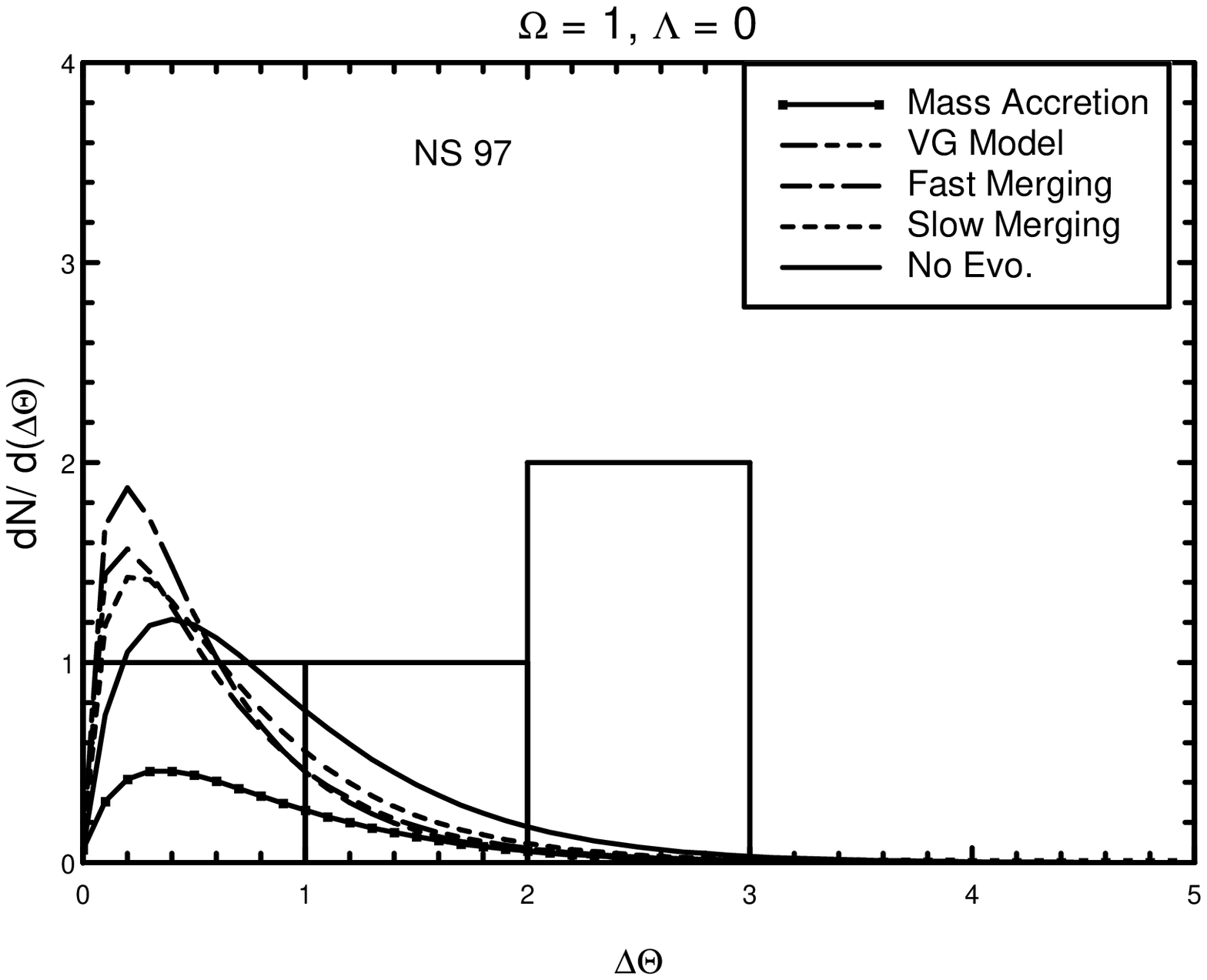}
\caption{}{ The expected distribution of lens image separations
with NS97 parameters with $\Omega = 1.0, \Lambda = 0.0$}
\end{figure}

\end{document}